\documentclass[12pt]{iopart}

\usepackage{iopams}  
\usepackage{graphicx}
\begin{document}

\title[Compensation of the Kondo effect in quantum dots coupled to 
       ferromagnetic leads ...]
       {Compensation of the Kondo effect in quantum dots coupled to 
       ferromagnetic leads within equation of motion approach}

\author{Mariusz Krawiec}
\address{Institute of Physics and Nanotechnology Center, 
         M. Curie-Sk\l odowska University, Pl. M. Curie-Sk\l odowskiej 1,
	 20-031 Lublin, Poland}
\ead{krawiec@kft.umcs.lublin.pl}

\begin{abstract}
We propose a new approximation scheme within equation of motion approach (EOM) 
to spin polarized transport through a quantum dot coupled to ferromagnetic 
leads. It has some advantages over a widely used in the literature standard EOM 
technique, in particular when we are interested in spin polarized quantities. 
Namely, it gives the values of the dot spin polarization which are closer to 
the ones obtained within numerical renormalization group (NRG), than the 
standard EOM approach. While restoring the Kondo effect, the spin polarization 
vanishes and the transport becomes unpolarized, in agreement with NRG and a 
real time diagrammatic calculations. The standard EOM procedure gives nonzero 
values of the spin polarization, and the transport is still spin polarized. 
Both approximations give the same correct splitting of the Kondo peaks due to 
ferromagnetism in the electrodes. 
\end{abstract}

\maketitle


\section{\label{Introduction} Introduction}

The growing experimental interest in magnetic materials has evolved into a new
research field, spintronics, where the transport properties are governed by the 
electron spin rather than charge. Thus, it opened new possibilities for 
technological progress in nanoelectronics \cite{Prinz,Awschalom} and quantum
computing \cite{Awschalom,Loss}. On the other hand, due to continuing 
experimental progress in miniaturization of electronic devices it became 
possible to study the fundamental problems of quantum mechanics. One of such 
examples is the Kondo effect \cite{Hewson} in quantum dots (QD). 

The Kondo effect can be observed when the dot has unpaired spin or, in other
words, has odd number of electrons. Thus the unpaired spin on the dot forms a 
many body singlet state with conduction electron spins in the leads. This state
manifests itself in a resonance at the Fermi energy in the dot density of 
states and zero-bias maximum in differential conductance. The Kondo effect was
predicted a long ago \cite{Glazman,Ng,Kawabata}, extensively studied
theoretically \cite{Meir,Herschfield,Wingreen,MK_1} and confirmed in a series 
of beautiful experiments \cite{Goldhaber,Cronenwett,Schmid,Simmel,Sasaki} in 
QD coupled to normal (non-magnetic) leads.

If the normal leads are replaced by ferromagnetic ones, spin degrees of freedom
start to play significant role modifying transport
\cite{Pasupathy,Nygard,Sergueev,Martinek_1,Lu,Zhang,Ma_1,Bulka,Lopez,Dong_1,
Martinek_2,Choi,Ma_2,Konig,Martinek_3,Tanaka,Sanchez,Utsumi,Swirkowicz} and 
thermoelectric properties \cite{MK_2}, eventually leading to new phenomena. One
of such new effects, associated with the ferromagnetism in the leads, is a 
splitting of the Kondo resonance due to spin dependent quantum charge 
fluctuations 
\cite{Martinek_1,Dong_1,Martinek_2,Choi,Sanchez,Utsumi,Swirkowicz,MK_2}. This
splitting depends on the magnitude of the leads magnetizations as well as on
their alignment in both leads. In particular, when the leads magnetizations
point in opposite directions (anti-parallel alignment), one observes no
splitting of the resonance and full Kondo effect is present for all spin
polarizations. However while measuring differential conductance the Kondo
resonance gets suppressed with increasing of the leads polarization, finally 
leading to a complete disappearance of zero-bias anomaly for fully polarized 
electrodes (p=1). In this case transport is completely blocked and no current 
flows for any voltage. On the other hand, in parallel alignment, the Kondo 
resonance is split and also gets suppressed with increasing the leads 
polarization. However, in this case transport is not completely blocked even 
for p=1, as one spin channel is still conducting \cite{MK_3}. 

Another interesting phenomenon is a compensation of the Kondo effect by 
external magnetic field ($B$). As it was mentioned before, at zero magnetic 
field in parallel configuration, one observes splitting of the Kondo resonance. 
Moreover, finite spin polarization is induced on the dot due to ferromagnetic 
electrodes. It turns out, that applying external magnetic field one can recover 
full Kondo effect, i.e. no splitting of the zero-energy resonance and vanishing 
of spin polarization on the dot. The transport becomes unpolarized. Thus, at 
certain magnetic field $B = B_{comp}$, which we call it compensating field, the 
strong coupling limit is recovered. However, the problem is with standard 
equation of motion approach (EOM) \cite{Martinek_1,MK_2}, as it gives non-zero 
spin polarization and spin polarized conductance, even at $B = B_{comp}$, i.e. 
when there is no splitting of the Kondo resonance. This is in contradiction 
with other approaches, like numerical renormalization group (NRG) 
\cite{Martinek_2} and a real-time diagrammatic technique \cite{Utsumi}, which 
correctly give zero spin polarization and equal spin dependent contributions to
the conductance at $B = B_{comp}$. It is the purpose of the present paper to 
show how one can cure a standard EOM approach from their disabilities of 
non-zero spin polarization and spin polarized conductance at $B = B_{comp}$.

The paper is organized as follows: in Sec. \ref{Model} the model and details of
calculations are presented. Section \ref{Density} shows a comparison of density
of states and the on-dot occupations, obtained in different approaches. Section 
\ref{Recovery} is devoted to compensation of the Kondo effect within the 
standard and the present EOM approaches, and finally conclusions are given in 
Sec. \ref{Conclusions}.


\section{\label{Model} Model and formulation}

Our model system, i.e. quantum dot coupled to external leads is represented by
a single impurity Anderson model in the limit of strong on-dot Coulomb 
repulsion ($U \rightarrow \infty$). The model Hamiltonian in the slave boson
representation \cite{LeGuillou,MK_2,MK_4,MK_5} reads
\begin{eqnarray}
H = \sum_{\lambda {\bf k} \sigma} \epsilon_{\lambda {\bf k} \sigma} 
    c^+_{\lambda {\bf k} \sigma} c_{\lambda {\bf k} \sigma} +
    \sum_{\sigma} \varepsilon_{\sigma} f^+_{\sigma} f_{\sigma} +
\nonumber \\
    \sum_{\lambda {\bf k}} \left(V_{\lambda {\bf k} \sigma} 
    c^+_{\lambda {\bf k} \sigma} b^+ f_{\sigma} + H. c. \right),
\label{Hamilt}
\end{eqnarray}
where $\lambda = L$ ($R$) denotes left (right) lead, 
$c^+_{\lambda {\bf k} \sigma}$ ($c_{\lambda {\bf k} \sigma}$) is the creation 
(annihilation) operator for a conduction electron with the wave vector 
${\bf k}$, spin $\sigma$ in the lead $\lambda$, $f^+_{\sigma}$ ($f_{\sigma}$) 
is a fermion operator, creating (annihilating) spin $\sigma$ on the dot, while 
$b^+$ ($b$) is a boson operator responsible for creating (annihilating) an 
empty dot state. The product of the fermion and boson operators gives a real 
dot electron operator ($d_{\sigma} = b^+ f_{\sigma}$). 
$V_{\lambda {\bf k} \sigma}$ is the hybridization between localized electron on 
the dot with the energy $\varepsilon_{\sigma}$ and conduction electron of 
energy $\epsilon_{\lambda {\bf k}}$ in the lead $\lambda$. Ferromagnetism of 
the electrodes is modeled via spin dependent conduction energy bandwidths. The 
constraint of no double occupancy is exactly taken into account by the 
non-canonical commutation rules for fermion and boson operators 
\cite{LeGuillou}.

The total current $I = \sum_{\sigma} I_{\sigma}$, flowing through a quantum dot 
is given in the standard form \cite{Haug}
\begin{eqnarray}
I = \frac{e}{\hbar} \sum_{\sigma} \int d\omega 
\frac{\Gamma_{L \sigma}(\omega) \Gamma_{R \sigma}(\omega)}
{\Gamma_{L \sigma}(\omega) + \Gamma_{R \sigma}(\omega)} 
[f_L(\omega) - f_R(\omega)] \rho_{\sigma}(\omega),
\label{current}
\end{eqnarray}
where $\Gamma_{\lambda \sigma}(\omega) = 
2 \pi \sum_{\bf k} |V_{\lambda {\bf k}\sigma}|^2 
\delta(\omega - \epsilon_{\lambda {\bf k} \sigma})$ is the coupling parameter, 
$\rho_{\sigma}(\omega)$ is the on-dot spin dependent density of states, and 
$f_{\lambda}(\omega) = f(\omega - \mu_{\lambda})$ is the Fermi distribution 
function in the lead $\lambda$ with the chemical potential $\mu_{\lambda}$ and 
temperature $T$.

In order to get the density of states $\rho_{\sigma}(\omega)$ one has to
calculate on-dot retarded Green function (GF) $G^r_{\sigma}(\omega)$. Within 
equation of motion approach the resulted GF is
\cite{Suhl,Nagaoka,Appelbaum,Lacroix,Meir,Monreal}
\begin{eqnarray}
G^r_{\sigma}(\omega) = \frac{1 - \langle n_{-\sigma} \rangle}{\omega -
\varepsilon_{\sigma} - \Sigma_{0 \sigma}(\omega) - \Sigma_{I \sigma}(\omega)}
\label{Green}
\end{eqnarray}
with non-interacting ($U = 0$) 
\begin{eqnarray}
\Sigma_{0\sigma}(\omega) = 
\sum_{\lambda {\bf k}} \frac{|V_{\lambda {\bf k}}|^2}
{\omega - \epsilon_{\lambda {\bf k} \sigma} +i0^+} 
\label{Sigma_0}
\end{eqnarray}
and interacting self-energy 
\begin{eqnarray}
\Sigma_{I\sigma}(\omega) = \sum_{\lambda {\bf k}} 
\frac{|V_{\lambda {\bf k}}|^2 f_{\lambda}(\epsilon_{\lambda {\bf k} -\sigma})}
{\omega - \epsilon_{\lambda {\bf k} -\sigma} - \varepsilon_{-\sigma} + 
\varepsilon_{\sigma} +i0^+} ,
\label{Sigma_I}
\end{eqnarray}
which is responsible for the generation of the Kondo effect. The 
non-interacting self-energy $\Sigma_{0\sigma}(\omega)$ is an exact solution of 
the problem with no Coulomb interactions present. The interacting self-energy 
$\Sigma_{I\sigma}(\omega)$ is obtained by neglecting terms in the equation of
motion for $G^r_{\sigma}(\omega)$ which cannot be directly projected onto 
original dot GF at this stage \cite{high_EOM}. 

In the standard EOM approach the ferromagnetism in the leads is modeled via 
spin dependent coupling parameters $\Gamma_{\lambda \sigma}(\omega) = 
2 \pi |V^2_{\lambda}| \rho_{\lambda\sigma}(E_F)$, where $E_F$ is the Fermi 
energy in the lead $\lambda$. In order to get the splitting of the Kondo 
resonance due to ferromagnetism in the leads, one replaces bare dot energy 
level $\varepsilon_{\sigma}$ in self-energy $\Sigma_{I\sigma}(\omega)$ 
(Eq. \ref{Sigma_I}) by renormalized one $\tilde \varepsilon_{\sigma}$, 
self-consistently found from the relation \cite{Martinek_1}
\begin{eqnarray}
\tilde \varepsilon_{\sigma} = \varepsilon_{\sigma} + 
{\it Re} [\Sigma_{0\sigma}(\tilde \varepsilon_{\sigma}) + 
\Sigma_{I \sigma}(\tilde \varepsilon_{\sigma})] .
\label{e_sigma}
\end{eqnarray}
In this way obtained splitting of the zero energy resonance remains in good 
agreement with a pour man scaling approach \cite{Martinek_1}. However such a
procedure has very important drawback, when the Kondo effect is compensated by 
external magnetic field $B$. While at $B = B_{comp}$ there is no splitting of 
the Kondo resonance, as it follows from other approaches 
\cite{Martinek_1,Martinek_2,Utsumi}, it gives non-zero value of on-dot spin 
polarization and not equal spin polarized contributions to the transport, 
which is in contradiction with NRG \cite{Martinek_2} and the real time 
diagrammatic calculations \cite{Utsumi}.

To see why the standard EOM approach gives non-zero spin polarization at 
$B = B_{comp}$ let us examine the structure of the dot GF (Eq. (\ref{Green})).
At zero magnetic field 
($\varepsilon_{\uparrow} = \varepsilon_{\downarrow} = \varepsilon_0$) 
the denominator of GF can be written in the form
\begin{eqnarray}
\omega - \varepsilon_0 - (1+p)\Sigma_0(\omega) - 
(1+p)\Sigma_I(\omega + \Delta \tilde \varepsilon)
\label{GF1up}
\end{eqnarray}
for spin up electrons, and 
\begin{eqnarray}
\omega - \varepsilon_0 - (1-p)\Sigma_0(\omega) - 
(1-p)\Sigma_I(\omega - \Delta \tilde \varepsilon)
\label{GF1down}
\end{eqnarray}
for spin down electrons, respectively. Both $\Sigma_0(\omega)$ and 
$\Sigma_I(\omega)$ contain spin independent couplings to the leads, and the 
spin dependence is shifted to the polarization parameter $p$, and 
$\Delta \tilde \varepsilon = 
\tilde \varepsilon_{\uparrow} - \tilde \varepsilon_{\downarrow}$ is calculated
form Eq. (\ref{e_sigma}). At $B = B_{comp}$ those equations are
\begin{eqnarray}
\omega - \varepsilon_{\uparrow} - (1+p)\Sigma_0(\omega) - 
(1+p)\Sigma_I(\omega)
\label{GF2up}
\end{eqnarray}
\begin{eqnarray}
\omega - \varepsilon_{\downarrow} - (1-p)\Sigma_0(\omega) - 
(1-p)\Sigma_I(\omega)
\label{GF2down}
\end{eqnarray}
where $\varepsilon_{\sigma} = \varepsilon_0 + \sigma B_{comp}$, and 
$\sigma \pm 1$. As one can notice, the real parts of both equations give the 
same values, that means the DOS for both spin directions will have a charge 
fluctuation resonance centered around the same energy. So in fact it should 
give the same occupations $n_{\uparrow} = n_{\downarrow}$. However, this is 
not the case, as we have to consider also the imaginary part of self-energies 
$\Sigma_0(\omega)$ and $\Sigma_I(\omega)$, which both depend on the 
polarization $p$. This leads to the smaller spin up occupation $n_{\uparrow}$ 
due to larger broadening of the charge fluctuation resonance in the DOS by 
factor $1 + p$. Similarly, $n_{\downarrow}$ is larger due to a factor $1 - p$ 
in the imaginary part of self-energies. As a result, spin polarization is 
non-zero at $B = B_{comp}$.  

To cure those inconsistencies of the standard EOM technique, we propose 
modifications of the approach. We start from two requirements: (i) at zero 
magnetic field it should give the same splitting of the Kondo resonance as the
standard EOM approach, and (ii) at $B = B_{comp}$ it should lead to zero spin
polarization and no splitting of the Kondo resonance. This could be easily
obtained if we skipped $1 \pm p$ factors in the imaginary parts of 
self-energies but left them in real parts. However such a procedure seems to be 
difficult to substantiate. Here, we propose slightly different approach which, 
by the construction, fulfills the above requirements. Namely, we use the same 
values of 
$\Gamma_{\lambda \uparrow} = \Gamma_{\lambda \downarrow} = \Gamma_{\lambda 0}$ 
for both spin directions but different spin dependent bandwidths 
$D_{\lambda \sigma}$ in the electrodes, which leads to spin asymmetry in the
electrodes. This is one of the ways of modeling a ferromagnetism in  the
electrodes, closely related to the Stoner model, but not unique 
\cite{Martinek_2}. The splitting of the Kondo resonance due to ferromagnetic 
leads 
($\Delta \tilde \varepsilon = 
\tilde \varepsilon_{\uparrow} - \tilde \varepsilon_{\downarrow}$) 
is obtained in the same way as in standard EOM approach (Eq. \ref{e_sigma}) 
with replaced $\Gamma_{\lambda \sigma}$ in $\Sigma_{0 \sigma}$ and 
$\Sigma_{I \sigma}$ by 
$\tilde \Gamma_{\lambda \sigma} = (1 + \sigma p_{\lambda}) \Gamma_{\lambda 0}$, 
where $p_{\lambda}$ is the polarization in the lead $\lambda$, and 
$\sigma = \pm 1$. Note that we use the same bare $\Gamma_{\lambda 0}$ for both 
spin directions in expression for the GF (Eq. (\ref{Green})). The polarization 
$p$ in the electrodes is calculated from spin dependent electron concentrations 
in the leads, i.e. $p = N_{\uparrow} - N_{\downarrow}$. Those concentrations 
will be different due to unequal bandwidths for spin up and spin down 
electrons. In this way obtained splitting of the zero energy resonance is the 
same as that obtained in pour man scaling and standard EOM technique 
\cite{Martinek_1}. Thus the requirement (i) is fulfilled. 

To fulfill the requirement (ii) let us write down the denominator of the dot 
GF (Eq. (\ref{Green})), similarly as for standard EOM equations. At zero
magnetic field it gives
\begin{eqnarray}
\omega - \varepsilon_0 - \Sigma_0(\omega) - 
\Sigma_I(\omega + \Delta \tilde \varepsilon)
\label{GF3up}
\end{eqnarray}
for spin up electrons, and 
\begin{eqnarray}
\omega - \varepsilon_0 - \Sigma_0(\omega) - 
\Sigma_I(\omega - \Delta \tilde \varepsilon)
\label{GF3down}
\end{eqnarray}
for spin down electrons, respectively. At $B = B_{comp}$ the above equations
read
\begin{eqnarray}
\omega - \varepsilon_{\uparrow} - \Sigma_0(\omega) - \Sigma_I(\omega)
\label{GF4up}
\end{eqnarray}
\begin{eqnarray}
\omega - \varepsilon_{\downarrow} - \Sigma_0(\omega) - \Sigma_I(\omega)
\label{GF4down}
\end{eqnarray}
where again $\varepsilon_{\sigma} = \varepsilon_0 + \sigma B_{comp}$. As we can 
see, there is no splitting of the Kondo resonance, however, it also leads to 
non-zero spin polarization due to different real parts of the above equations. 
In this case the charge fluctuation resonances in spin polarized DOS are 
centered at different energies. Perhaps the easiest way of achieving zero spin
polarization at $B = B_{comp}$ is to replace the original dot energies 
$\varepsilon_{\sigma} = \varepsilon_0 + \sigma B$ in the above equations by 
the same energies $\varepsilon_{\sigma}$ but expressed in terms of 
$\tilde \varepsilon_{\sigma}$ and 
${\it Re} \Sigma_I(\tilde \varepsilon_{\sigma})$, calculated from 
Eq. (\ref{e_sigma}). Thus at $B = 0$ we get splitting of the Kondo resonance 
and non-zero spin polarization, while at $B = B_{comp}$ the splitting vanishes 
as well as the spin polarization. However, such a procedure is equivalent to 
leaving the original dot energy level $\varepsilon_0$  unchanged, even in the 
presence of the external magnetic field. In other words, we have to assume that 
the $B$ field modifies $\Sigma_I(\omega)$ only - there is no Zeeman splitting 
of the dot energy level. 

Here we propose another, a more natural way of getting the zero spin 
polarization at $B = B_{comp}$. Namely, we replace the dot energies 
$\varepsilon_{\sigma}$ by $\tilde \varepsilon_{\sigma}$, not only in the 
interacting self-energy $\Sigma_{I \sigma}$ (Eq. \ref{Sigma_I}), as in standard 
EOM, but also in full retarded Green function $G^r_{\sigma}$ (Eq. \ref{Green}). 
This corresponds to the same replacement of the energies in Eqs. (\ref{GF4up}) 
and (\ref{GF4down}), which in turn leads to the same expressions for Green 
functions for both spin directions at $B = B_{comp}$. Such a renormalization of 
$\varepsilon_{\sigma}$, in a heuristic way, represents fact that higher order 
GFs in EOM procedure give also $V^2$ contributions (similar to 
$\Sigma_{I\sigma}(\omega)$) going like $ln(\omega)$ near the Fermi energy 
\cite{Monreal}, which are neglected in standard $V^2$ EOM approach. Perhaps 
this is a simplest way of enhancing spin correlations, i.e. a transfer of the 
spectral weight from charge to spin sector by the renormalization of the dot 
energy level. This in turn leads to better description of the Kondo effect, as 
it will be shown later on (see discussion in Sec. \ref{Density}). Such a 
procedure gives the correct splitting of the Kondo resonance 
$\Delta \tilde \varepsilon$, equal spin dependent contributions to the 
conductance as well as the vanishing of the spin polarization at 
$B = B_{comp}$. Thus both requirements (i) and (ii) are automatically 
fulfilled. Of course, such a modification does not cure all the shortcomings of 
standard EOM approach, like basic Fermi liquid relations, which are little bit
less but still violated. However, it leads to better qualitative description of 
the Kondo effect in quantum dots.

In the following we show how those modifications of the approach influence the
properties of the quantum dot in the presence of ferromagnetism. In numerical 
calculations all the energies are measured with respect to the Fermi energy 
$E_F = 0$ in units of $\Gamma = \Gamma_{L0} + \Gamma_{R0} = 1$.


\section{\label{Density} Density of states}

The most influenced quantity by the above modifications of the EOM approach is
the density of states (DOS) 
$A_{\sigma}(\omega) = -\frac{1}{\pi} G^r_{\sigma}(\omega)$. 

\begin{figure}[h]
 \begin{center}
 \resizebox{0.5\linewidth}{!}{
  \includegraphics{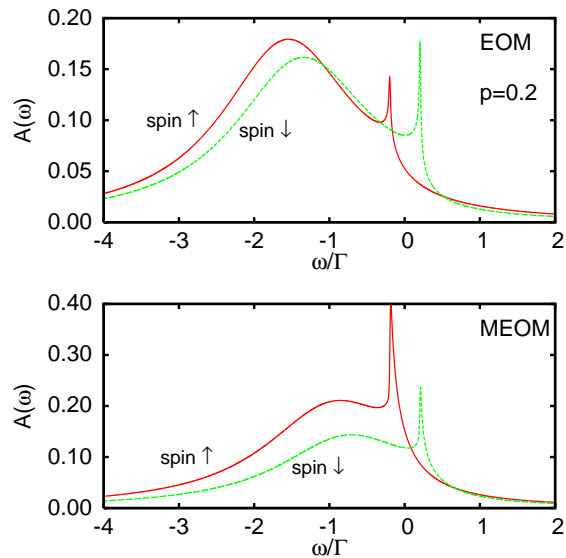}}
 \end{center}
 \caption{\label{Fig1} Comparison of spin up (solid line) and 
          spin down (dashed line) dot density of states obtained in the 
	  standard EOM (top panel) and in the present approach (bottom panel). 
	  The model parameters are: T=0.005, $\varepsilon_{\sigma}$=-2 in units 
	  of $\Gamma$. The polarization of the electrodes is p$_L$=p$_R$=0.2.}
\end{figure}

Figure \ref{Fig1} shows a comparison of spin dependent densities of states 
obtained in the standard EOM (top panel) and in the present, modified EOM
approach (bottom panel).
First of all, one can see that the positions of the Kondo resonances remain 
unchanged in both approaches, giving the same value of the splitting. However,
there is a change in their spectral weights. While in the standard EOM the 
Kondo resonance for spin down electrons is higher, in the modified equation of
motion technique (MEOM) the resonance for spin up electrons is more pronounced, 
similarly as in real time diagrammatic calculations \cite{Utsumi}. Moreover, 
the DOS around the dot energy level $\varepsilon_{\sigma}$ also changes leading 
to higher values for spin up electrons. This is also in agreement with real 
time diagrammatic approach (see Fig. 6 of Ref. \cite{Utsumi}). At the same time 
the standard EOM gives comparable or only slightly higher DOS around 
$\varepsilon_{\sigma}$. At the first sight, the higher Kondo resonance for spin
up electrons seems to be counterintuitive, as there is more spin up electrons 
in the leads, thus they should lead to better screening of spin down electrons 
on the dot. However, one has to remember that due to spin dependent 
renormalization of the dot energy level, the spin up electron occupation on the
dot is larger, thus the cotunneling processes, including Kondo ones, are more
efficient in spin up channel.

\begin{figure}[h]
 \begin{center}
 \resizebox{0.5\linewidth}{!}{
  \includegraphics{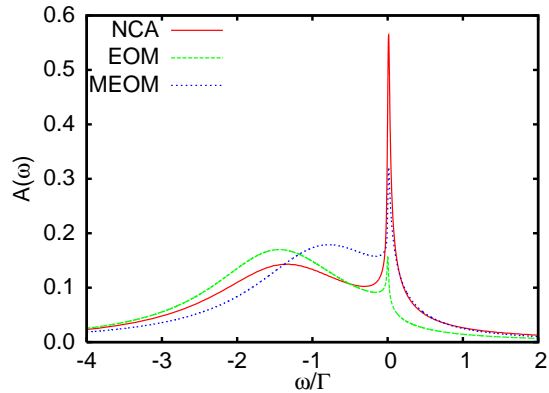}}
 \end{center}
 \caption{\label{Fig2} The dot density of states obtained in NCA 
          (solid line), standard EOM (dashed line) and modified EOM (dotted 
	  line) approach. The model parameters are the same as in Fig. 
	  \ref{Fig1} but the electrodes are unpolarized now (p=0).}
\end{figure}

As one can notice, the MEOM approach seems to give to large renormalization of 
the dot energy level $\varepsilon_{\sigma}$, shifting the charge fluctuation 
resonance in the DOS towards the Fermi energy for both spin directions (compare 
both panels of Fig. \ref{Fig1}). However such a renormalization influences 
also the DOS around $E_F$ leading to better description of the Kondo resonance.
This can be seen in Fig. \ref{Fig2}, where the densities of states for 
unpolarized leads (p=0) obtained in different approaches are shown. 
The solid line is obtained within non-crossing approximation (NCA) 
\cite{Keiter}, which is a widely accepted and reliable technique for 
description of the Kondo problem in the case of non-magnetic leads and in the
lack of external magnetic field \cite{Wingreen,Costi,Hettler,MK_1}, the dashed 
one within standard EOM, while the dotted one within the present approach. 
Clearly, the present approach gives better behavior of the DOS in the low 
energy regime, thus better description of the Kondo effect. 

Physically, as it was previously mentioned, the renormalization of
$\varepsilon_{\sigma}$ in GF represents $V^2$ contributions to GF, which are
obtained while calculating higher order GFs in EOM procedure. In particular,
performing an effective $V^4$ order EOM calculations, one gets $V^2$ 
contributions, going like $ln(\omega)$ around the Fermi energy \cite{Monreal}.
Those contributions are very important for the Kondo effect, as they represent 
inelastic scattering processes leading to the broadening of the Kondo resonance 
(compare Fig. \ref{Fig2} and Fig. 1 of Ref. \cite{Monreal}). Here similar 
effect is achieved simply just by renormalization of the dot energy level, 
which in turn leads to a transfer of the spectral weight form charge to spin 
sector enhancing the Kondo effect. Furthermore, this can be qualitatively 
explained even within the EOM approach. Namely, while calculating original QD 
Green function one obtains, usually neglected, an energy independent term 
$ \Lambda = - \sum_{\lambda {\bf k}} V_{\lambda {\bf k}} 
\langle f^+_{-\sigma} b c_{\lambda {\bf k} -\sigma} \rangle$, which shifts the
QD bare energy level towards $E_F$. In equilibrium, the $\Lambda$ can be
expressed in terms of the dot GF, i.e. 
$\Lambda = \frac{1}{\pi} \sum_{\lambda {\bf k}} V^2_{\lambda {\bf k}} 
\int d\omega f(\omega) {\it Im} \left(
\frac{G^r_{-\sigma}(\omega)}{\omega - \epsilon_{\lambda {\bf k} -\sigma}} 
\right)$. Clearly, this is $V^2$ contribution renormalizing QD energy level 
in the same way as we propose here. This contribution has to be also calculated 
self-consistently, as it depends on the QD energy level, and explains why we
replaced original QD energy level $\varepsilon_{\sigma}$ by 
$\tilde \varepsilon_{\sigma}$ also in full retarded QD Green function
$G^r_{\sigma}(\omega)$ (Eq. (\ref{Green})).

Another important consequence associated with the above modifications of the 
EOM approach is the average dot occupation $n = n_{\uparrow} + n_{\downarrow}$.
For unpolarized leads and the dot energy level $\varepsilon_{\sigma} = -2$, the
MEOM gives $n_{MEOM} = 0.87$ which is close to value obtained within NCA 
($n_{NCA} = 0.85$), while the standard EOM deviates from $n_{NCA}$ by almost 
$10 \%$, giving $n_{EOM} = 0.93$.


\section{\label{Recovery} Recovery of the Kondo effect}

Now, let us discuss the effect of the external magnetic field on the Kondo 
effect. In the following we assume that the external magnetic field acts on 
the dot spin only and disregard its influence on the properties of the leads. 
In real experiment, this cannot be neglected, as it can lead to the 
modifications of the magnetic properties and the density of states in the 
electrodes \cite{Pustilnik}. However, it needs fully self-consistent 
calculations, which are out of the scope of the present work. 

It turns out that applying magnetic field one can recover the Kondo 
effect. At certain field $B = B_{comp}$ there is no splitting of the zero 
energy resonances (in real experiments the $B$ field will also modify DOS in 
the electrodes \cite{Pustilnik}), the transport is unpolarized, and the spin 
polarization should vanish in this case, so the strong coupling limit is 
reached. Thus, full Kondo effect is recovered. Those conclusions have been 
obtained within NRG \cite{Martinek_2} and a real time diagrammatic technique 
\cite{Utsumi}. Also standard EOM approach gives no splitting of the Kondo 
resonance in this case \cite{Martinek_1} but does not fulfill the conditions 
of vanishing spin polarization and equal spin dependent contributions to the 
conductance at $B_{comp}$. The proposed modifications of the standard EOM 
substantially improve the results leading also to unpolarized transport and 
to zero spin polarization at $B_{comp}$.

\begin{figure}[h]
 \begin{center}
 \resizebox{0.5\linewidth}{!}{
  \includegraphics{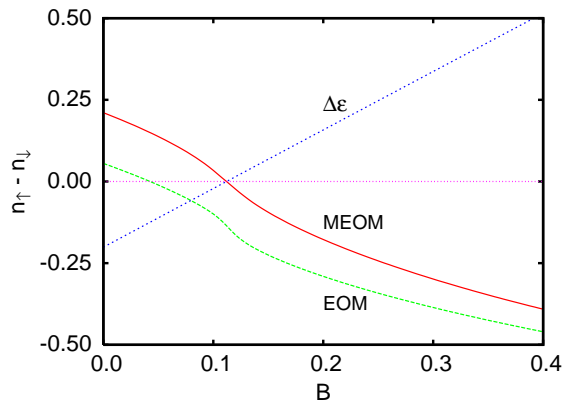}}
 \end{center}
 \caption{\label{Fig3} The spin polarization 
          $n_{\uparrow} - n_{\downarrow}$ calculated within the present 
	  approach (solid line) and within standard EOM (dashed line). Note 
	  that within MEOM approach spin polarization vanishes at 
	  $B = B_{comp} = 0.11$, while EOM gives non-zero value of it. The 
	  splitting of the zero energy resonance is also shown (dotted line). 
	  The model parameters are the same as in Fig. \ref{Fig1}.}
\end{figure}

Figure \ref{Fig3} shows a comparison of the spin polarization 
$n_{\uparrow} - n_{\downarrow}$ vs. magnetic field $B$, calculated within the
present approach (solid line) and the standard EOM technique (dashed line).
The dotted line represents the splitting of the zero energy resonance 
$\Delta \tilde \varepsilon_{\sigma}$. It is clearly seen that the spin 
polarization obtained within MEOM vanishes, while standard EOM gives non-zero 
value of it at $B = B_{comp} = 0.11$, for which 
$\Delta \tilde \varepsilon_{\sigma} = 0$. At this field standard EOM gives 
$n_{\uparrow} - n_{\downarrow} = -0.135$. The spin polarization vanishes at 
much smaller $B$ field, i.e. at $B = 0.042$.

\begin{figure}[h]
 \begin{center}
 \resizebox{0.5\linewidth}{!}{
  \includegraphics{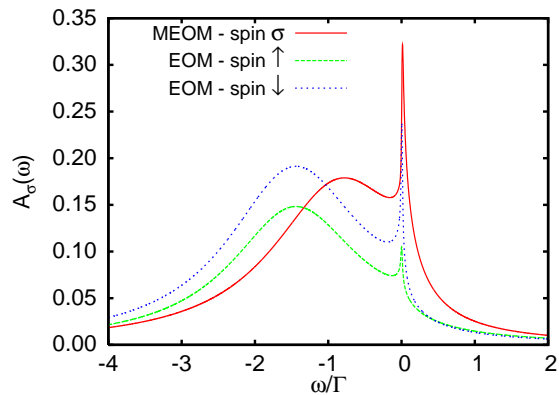}}
 \end{center}
 \caption{\label{Fig4} Spin dependent density of states at 
          $B = B_{comp}$. Modified EOM gives exactly the same DOS for both 
	  spin directions (solid line), while the standard EOM gives different 
	  densities of states for spin up (dashed line) and spin down electrons 
	  (dotted line). The parameters are the same as previously used.}
\end{figure}

The non-zero spin polarization results form different densities of states for
spin up and spin down electrons, which can be seen in Fig. \ref{Fig4} (dashed
and dotted line) or Fig. 1(d) of Ref. \cite{Martinek_1}.
Clearly, the spin down DOS is much larger than spin up one, even if there is no
splitting of the Kondo resonance. Similarly, other approaches, like NRG or 
real time diagrammatic approach, give also asymmetric density of states but at
the same time they give zero spin polarization. This is achieved in the
following way. Spin down electrons have larger DOS at the Fermi energy but
smaller around $\varepsilon_{\sigma}$, in comparison with spin up electrons, so
the resulting DOS integrated with the Fermi distribution functions give the 
same occupations, thus no spin polarization \cite{Martinek_2,Utsumi}. The MEOM 
approach also gives zero spin polarization but the price we have to pay for 
this are the same densities of states for both spin directions (see solid line 
in Fig. \ref{Fig4}). 

\begin{figure}[h]
 \begin{center}
 \resizebox{0.6\linewidth}{!}{
  \includegraphics{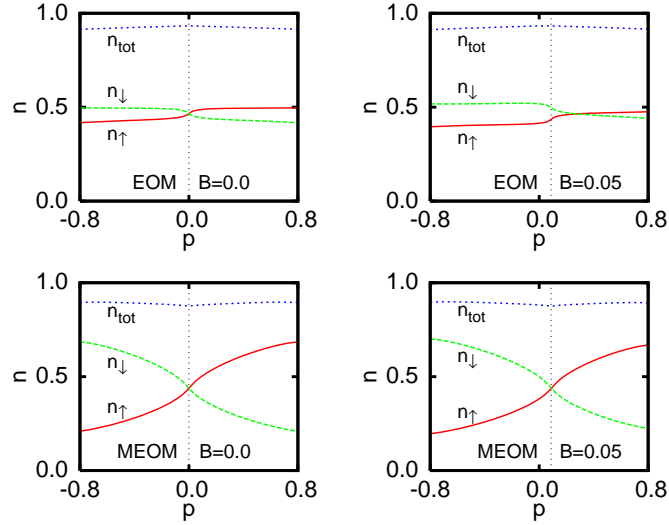}}
 \end{center}
 \caption{\label{Fig5} Spin-dependent occupation of the dot at 
          $B = 0$ (left panels) and $B = 0.1$ (right panels) as a function of 
	  spin polarization p. Top panels represent results obtained within 
	  standard EOM, while bottom ones - within the present approach. At 
	  $B =0$ both approaches give $n_{\uparrow} = n_{\downarrow}$ for 
	  p=0. At $B=0.05$ MEOM gives $n_{\uparrow} = n_{\downarrow}$ for a
	  finite p, for which $\Delta \tilde \varepsilon_{\sigma} = 0$, while 
	  the EOM gives $n_{\uparrow} \neq n_{\downarrow}$ in this case.}
\end{figure}

It is interesting to see the behavior of the dot spin-dependent occupations as 
a function of the leads polarization in both approaches. Corresponding spin up,
spin down and the total occupations are shown in Fig. \ref{Fig5} at zero
magnetic filed (left panels) and at $B = 0.05$ (right panels).
The top panels represent occupations obtained with help of standard EOM, while 
bottom panels - those obtained by the present, modified EOM approach. As one 
can see, at $B = 0$ both approaches give $n_{\uparrow} = n_{\downarrow}$ for 
unpolarized leads (p=0). However, the difference between $n_{\uparrow}$ and 
$n_{\downarrow}$, (spin polarization) obtained in the standard EOM is much 
smaller than in the modified EOM approach. Unfortunately, both approaches give
too small values of spin polarizations in comparison to the other approaches  
\cite{Martinek_2,Utsumi}. At finite magnetic field, modified EOM gives 
$n_{\uparrow} = n_{\downarrow}$ for a finite p, for which 
$\Delta \tilde \varepsilon_{\sigma} = 0$. On the other hand, the standard EOM 
gives $n_{\uparrow} \neq n_{\downarrow}$, in contradiction to NRG results 
(compare Fig. 1 of Ref. \cite{Martinek_2}).

While the standard EOM approach gives worse (smaller) values of spin 
polarization, it gives better behavior of the total occupation 
($n_{\uparrow} + n_{\downarrow}$) vs. leads polarization. In the standard EOM 
approach the total occupation weakly increases with decreasing of the leads
polarization p (top panels of Fig. \ref{Fig5}), surprisingly giving a correct 
position of the maximum of the total occupation, in agreement with NRG results 
\cite{Martinek_2}. In the present approach, the situation is opposite, namely, 
the total occupation weakly decreases with decreasing of the leads 
polarization, giving a correct position of a minimum, not a maximum of the 
total occupation. In both cases, however, the changes of the total occupations
due to the leads polarizations are very small.

\begin{figure}[h]
 \begin{center}
 \resizebox{0.5\linewidth}{!}{
  \includegraphics{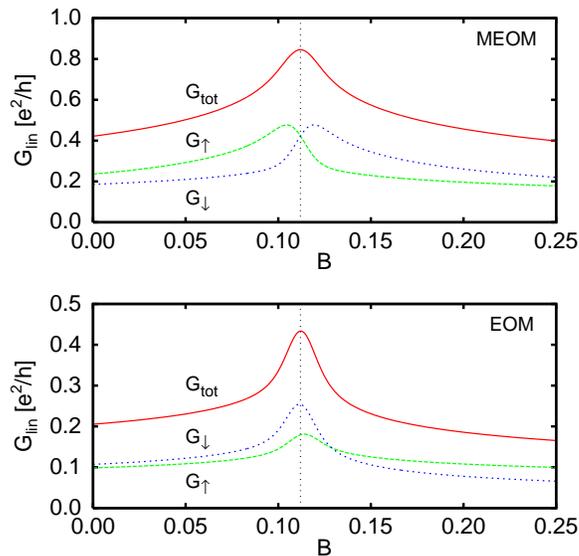}}
 \end{center}
 \caption{\label{Fig6} The linear conductance $G_{lin}$ (solid
          lines) and spin polarized contributions to it (dashed and dotted 
	  lines) vs. magnetic field $B$, calculated within the present approach 
	  (top panel) and the standard EOM technique (bottom panel). Note 
	  different behavior of $G_{\sigma}$ in both approaches.}
\end{figure}

Finally, let us turn our attention to the transport properties. Figure
\ref{Fig6} shows a linear conductance 
$G_{lin} = \frac{dI}{d(eV)}|_{eV \rightarrow 0}$ vs. external magnetic field 
$B$, calculated within the present approach (top panel) and the standard EOM
technique (bottom panel).
While both approaches lead to similar behavior of the total conductance
$G_{lin}$, i.e. a maximum at $B = B_{comp} = 0.11$ (indicated by thin vertical 
line), they give different behavior of spin polarized contributions 
$G_{\sigma}$ to it. Within the MEOM approach (top panel) the transport is
governed by majority spin electrons at small $B$ fields ($B < B_{comp}$), and 
by minority spin electrons at higher fields ($B > B_{comp}$). Such asymmetry of
$G_{\sigma}$ can be explained by the spin state of QD, similarly as the
asymmetry of the DOS (see Fig. \ref{Fig1}). The peak in $G_{\uparrow}$ (dashed
line) steams from the fact that at $B < B_{comp}$ the QD is occupied by spin up
electrons ($n_{\uparrow} > n_{\downarrow}$) (see Fig. \ref{Fig3}), thus the spin
up component of the cotunneling current is dominant. At $B > B_{comp}$ the
situation is opposite, namely, $n_{\uparrow} < n_{\downarrow}$, thus spin down
cotunneling current is larger. At $B = B_{comp}$ both spin channels equally 
contribute to the transport, again, indicating that full Kondo effect is 
recovered in this case. Note that, the calculated spin polarization vanishes in 
this case (see Fig. \ref{Fig3}). Such a behavior of the conductance and the 
spin polarization remains in good agreement with a real time diagrammatic 
technique (compare Fig. 6 (a) of Ref. \cite{Utsumi}). Within the MEOM approach, 
the condition $G_{\uparrow} = G_{\downarrow}$ at $B = B_{comp}$ steams from the 
fact that the density of states for both spin directions are the same (see Fig. 
\ref{Fig4}) and the couplings to the leads are equal 
($\Gamma_{\lambda \uparrow} = \Gamma_{\lambda \downarrow} = 
\Gamma_{\lambda 0}$).

The situation is somewhat worrying in the standard EOM approach (bottom panel),
as it gives an opposite behavior of $G_{\sigma}$. At $B < B_{comp}$, larger 
contribution comes form minority electrons, and at $B > B_{comp}$, it comes 
from majority electrons. Moreover, at $B = B_{comp}$, the transport still is 
spin polarized ($G_{\uparrow} \neq G_{\downarrow}$). Accidentally, the 
transport becomes unpolarized at slightly larger field than $B_{comp}$, while
spin polarization vanishes at different filed $B < B_{comp}$ (compare Fig.
\ref{Fig3}).

\begin{figure}[h]
 \begin{center}
 \resizebox{0.5\linewidth}{!}{
  \includegraphics{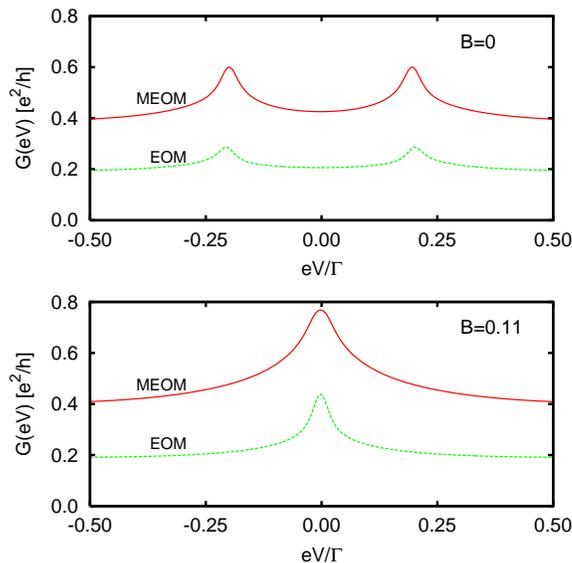}}
 \end{center}
 \caption{\label{Fig7} Differential conductance 
          $G(eV) = \frac{dI}{d(eV)}$ vs. bias voltage $eV = \mu_L - \mu_R$ of 
	  a quantum dot coupled to ferromagnetic leads with p=0.2. Top panel 
	  represents $G(eV)$ without external magnetic field, while the bottom 
	  one is for $B = B_{comp} = 0.11$. Solid lines are obtained with help 
	  of the present (MEOM) approach, and dashed lines - within standard 
	  EOM technique.}
\end{figure}

Under non-equilibrium conditions, the discrepancies between standard EOM and 
MEOM approaches look similar. Figure \ref{Fig7} shows a comparison of the 
differential conductance $G(eV) = \frac{dI}{d(eV)}$ vs. bias voltage 
$eV = \mu_L - \mu_R$ at zero magnetic field (top panel) and at 
$B = B_{comp} = 0.11$ (bottom panel), obtained within the present approach 
(solid lines) and the standard EOM scheme (dashed lines).
Both approaches give qualitatively similar behavior of the total conductance, 
i.e. the same splitting of the zero bias anomaly at zero magnetic field and its 
lack when the Kondo effect is compensated ($B = 0.11$). However, the present 
approach gives larger values of the conductance. Moreover, within the present 
approach (not shown here) both spin channels equally contribute to the transport 
at any voltage, when the Kondo effect is recovered by the external magnetic 
field $B = B_{comp}$. On the other hand, standard EOM technique leads to a spin
polarized $G(eV)$ (see Fig. 2 (d) of Ref. \cite{Martinek_1}). 

\begin{figure}[h]
 \begin{center}
 \resizebox{0.5\linewidth}{!}{
  \includegraphics{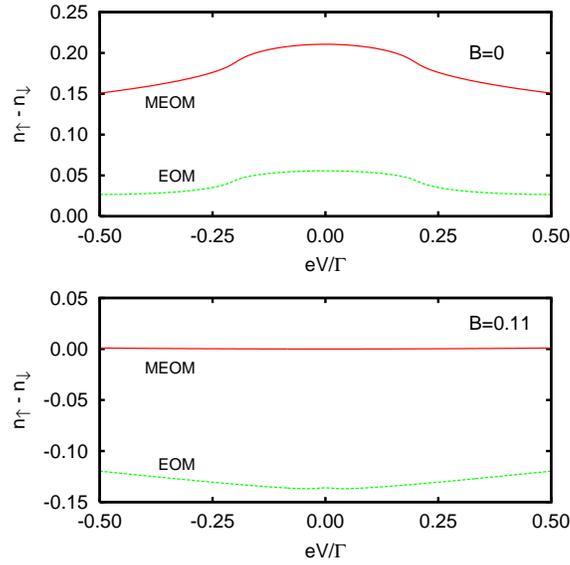}}
 \end{center}
 \caption{\label{Fig8} Spin polarization $n_{\uparrow} - n_{\downarrow}$ as a 
          function of applied bias voltage $eV$ in zero magnetic field (top 
	  panel) and at $B = B_{comp} = 0.11$ (bottom panel). Solid lines 
	  represent the results obtained by MEOM, and dashed lines - by 
	  standard EOM approach. Note that within the standard EOM spin 
	  polarization does not vanish at $B = B_{comp}$.}
\end{figure}

Corresponding voltage dependence of the spin polarization is shown in Fig.
\ref{Fig8}.
Again, at zero magnetic field (top panel), both approaches give qualitatively 
similar behavior of the dot spin polarization for not too large voltages, in 
agreement with a real time diagrammatic calculations \cite{Utsumi}. At higher 
voltages (not shown), in the standard EOM approach the spin polarization weakly 
increases with $eV$, while it still decreases in the present approach. On the
other hand, when the Kondo effect is compensated by the external $B$ field
(bottom panel), standard EOM approach gives non-zero values of the spin
polarization, while the modified one gives vanishing of it at $eV = 0$ and very
weak increasing with the voltage.

All the above results indicate that the present approach should better describe 
the compensation of the Kondo effect, as the transport (linear and non-linear
conductance) is unpolarized, there is no splitting of the Kondo resonance and 
spin polarization vanishes in this case, similarly as in other approaches 
\cite{Martinek_2,Utsumi}. On the other hand, standard EOM scheme also gives no
splitting of the Kondo resonance but at the same time it gives non-zero spin 
polarization and spin polarized transport properties.


\section{\label{Conclusions} Conclusions}

In conclusion we have proposed simple modifications of the standard equation of
motion approach to get a better description of the Kondo effect in a quantum 
dot coupled to ferromagnetic leads. Special emphasis was put to the 
compensation of the Kondo effect by external magnetic field. In particular, the 
present approach correctly gives both, no splitting of the Kondo resonance and 
the vanishing of the dot spin polarization at compensating magnetic field, 
while the standard equation of motion approach gives non-zero value of the
polarization. Moreover, the transport also remains unpolarized within the 
present approach, while the standard EOM gives not equal spin dependent 
contributions to the linear and non-linear conductance when the Kondo effect is
recovered. In spite of lack of reliable techniques for studying the 
non-equilibrium transport, the present approach could help us to understand 
the properties of a quantum dot in the presence of ferromagnatism in the
electrodes.

\section*{Acknowledgments}
This work has been supported by the Polish Ministry of Education and Science
under Grant No. N202 081 31/0372.


\end{document}